# Distributed Joint Offloading Decision and Resource Allocation for Multi-User Mobile Edge Computing: A Game Theory Approach

Ning Li , *Student Member*, *IEEE*, Jose-Fernan Martinez-Ortega, Gregorio Rubio

*Abstract-* Due to the spectrum reuse in small cell network, the inter-cell interference has great effection on MEC's performance. In this paper, for reducing the energy consumption and latency of MEC, we propose a game theory based jointing offloading decision and resource allocation algorithm for multi-user MEC. In this algorithm, the transmission power, offloading decision, and mobile user's CPU capability are determined jointly. We prove that this game is an exact potential game and the NE of this game exists and is unique. For reaching the NE, the best response dynamic is applied. We calculate the best responses of these three variables. Moreover, we investigate the properties of this algorithm, including the convergence, the computation complexity, and the price of anarchy. The theoretical analysis shows that the inter-cell interference has great effection on the performance of MEC. The NE of this game is Pareto efficiency and also the global optimal solution of the proposed optimal issue. Finally, we evaluate the performance of this algorithm by simulation. The simulation results illustrates that this algorithm is effective on improving the performance of the multi-user MEC system.

*Index Terms*-Mobile edge computing, offloading decision, transmission power control, game theory, interference.

## I. INTRODUCTION

With the development of smart mobile devices, more and more applications, which are computation-intensive, resource-hungry, and high energy consumption, emerge and attract great attention [1][2][3][4]. However, due to the computation capability and the energy of the mobile devices are always limited, the quality of computation experience cannot be satisfied. For releasing the tension between the resource limitation of the mobile devices and the requirements of the computation-intensive applications, the mobile edge computing (MEC) is proposed. Based on MEC, the requirements of low latency and energy-saving are satisfied by offloading the

Ning Li, Jose-Fernan Martinez-Ortega, and Gregorio Rubio are with the Universidad Politenica de Madrid, Madrid, Spain.
E-mail: {li.ning, jf.martinez, gregorio.rubio}@upm.es.

computation tasks to the resource-rich clouds that deployed by telecom operators [5]. The MEC can provide pervasive and agile computation augmenting services for mobile users at anytime and anywhere.

*A. Related works and Motivation*

The main purpose of MEC is to reduce energy consumption and latency by carefully designing the task offloading scheme and resource allocation manner [1][2][3][4], which has been learned separately or jointly in both the single-user and multi-user MEC system [6]-[18]. The resource allocation includes the channel allocation, CPU capability control, transmission power control, etc. In [6], the author proposes a game theory approach which can achieve efficient computation offloading for mobile cloud computing; the authors in [5] apply the conclusions in [6] into the multi-user MEC system and propose a game theory based offloading decision algorithm; in this algorithm, the mobile users decide whether offload the tasks to MEC servers and which communication channels are used in a distributed manner. In [7], the authors propose an optimal offloading policy for the applications which include sequential component dependency graphs and multi-radio enabled mobile devices. This policy can minimize the energy consumption of the mobile devices since the execution time is below a given threshold and the percentage of transmitted data is determined simultaneously. Different with the algorithm shown in [7], the authors in [8] solve the task offloading issues for arbitrary dependency graphics; in this algorithm, the wireless-aware scheduling of the application components and the offloading strategy are taken into account jointly. In [9], the authors propose an effective computation model for MEC system, which joints the computation and communication cooperation together to improve the system performance. The task offloading decision and the computational frequency scaling are learned jointly in [10]; in this paper, the latency and energy consumption are minimized by jointly optimizing the task allocation decision and CPU frequency of the mobile user. In [11], the computational speed and transmission power of mobile device and the offloading ratio are optimized jointly for reducing energy consumption and latency via dynamic voltage scaling technology. In [12], the transmission power control and offloading decision in MEC system under single-user scenario is investigated. As an improvement, the authors in [13] and [14] expand the conclusions in [12] to the multi-user MEC system and combine with the resource allocation to improve the performance of multi-user MEC system by Lyapunov optimization. Similar to [13], the authors in [15] proposed a centralized algorithm for multi-server multi-user MEC system, which joints the task offloading and resource allocation together; the jointing optimal issue is formulated as a mixed integer non-liner program problem in this algorithm. The service delay in MEC is reduced in [16] through virtual machine migration and transmission power control; the users transmit data to cloudlet in a round robin fashion and the service delay is reduced by controlling cloudlets' transmission power. In [17], an integrated framework for computation offloading and interference management in MEC system is proposed, in which, the offloading decision, physical resource block allocation, and MEC computation resource allocation are taken into account to improve the performance of MEC.

However, there are some limitations in the commonly adopted assumptions of the algorithms introduced above. Due to the spectrum reuse in small cell network, the inter-cell interference is critical to the quality of service [18][19][20]. If too many mobile users choose the same wireless channel to offload the computation task simultaneously, they may cause severe interference to each other, which would reduce the data rate and increase the energy consumption. This hence can lead to low energy efficiency and long latency. In [5]-[11], the inter-cell interference is not taken into account when determine the offloading decision of the mobile user. It has been shown that efficient power control can mitigate inter-cell interference successfully in small cell network [18][19][20][21], so some transmission power control based algorithms are proposed for MEC, such as [12]-[16]. Nevertheless, the [12] is proposed for single-user MEC system; the transmission power control algorithm proposed in [16] is mainly for the cloudlets rather than the mobile users; the algorithms proposed in [13], [14], and [15] are effective in multi-user MEC system, but they are centralized which are not always efficient in the distributed system, such as the wireless sensor network and the IoT applications. The distributed power control schemes outperform the centralized counterparts due to less information exchange and computations [22].

*B. Main Contributions*

Based on the issues aforementioned, we propose a game theory based algorithm for MEC, which joints the transmission power control, the offloading decision, and the CPU capability of mobile user together to improve the performance of MEC. The contributions of this paper are:

1. we joints the transmission power control, the offloading decision, and the CPU frequency control of the mobile user for the multi-user MEC system in a distributed manner; in this algorithm, the game theory, which is a powerful approach on solving the distributed issues in wireless network [2][23], is applied;

2. For reaching the Nash Equilibrium (NE) of this game, we introduce the best response dynamic into the iteration process; since there are three variables in the optimal function and they are relevant, we investigate the relationship between these variables and how to calculate the best responses of these variables; we also prove that the main factor affects the performance of MEC is the inter-cell interference;

3. We investigated the properties of this game; first, we prove that the game introduced in this paper is an exact potential game; second, we prove that the NE of this game exists and is unique; third, we proved that this game is convergence and reaching the NE of this game is Polynomial Local Search (PLS) complete based on the best response dynamic;

4. We investigate the properties of the proposed algorithm in detail; first, we analyze the computation complexity of this algorithm and prove that the algorithm can be finished in polynomial time; second, we investigate the price of anarchic (PoA) of this game in terms of the network computation overhead; to the PoA, we conclude that reducing the inter-cell interference can improve the performance of MEC effectively; finally, we prove that the NE of this game is Pareto efficiency and also a global optimal solution of the issues shown in (11).

## II. NETWORK MODEL AND PROBLEM STATEMENT

There are $N$ mobile users in the network, denoted as $\boldsymbol{N} = \{1, 2, \ldots, N\}$. The mobile users are divided into $L$ small cells and each small cell is served by a base station (BS), denoted as $\boldsymbol{L} = \{1, 2, \ldots, L\}$. The mobile user $n$ which is in small cell $l$ is expressed as $n_l$. Each user can only connect to one BS that is selected based on long-term channel quality measurement [20]. In this paper, the OFDMA is used for the upper link communication of the task offloading from the mobile user to MEC server [18][19][20]; the universal frequency reuse deployment in which every cell shares the whole bandwidth is applied [18][19][20]. In OFDMA, the available spectrum is divided into $K$ subchannels and the index of these subchannels is $\boldsymbol{K} = \{1, 2, \ldots, K\}$. In the MEC system, each mobile user has a computation intensive task, which can be calculated locally or offloaded to the MEC server through BS that deployed proximity to the user. There are two offloading models for mobile user [1]: binary offloading and partial offloading. In this paper, for unifying these two schemas, we introduce the offloading ratio into the offloading decision as that used in [9] and [11].

**Definition 1.** The offloading ratio $\lambda \in [0, 1]$ is defined as the ratio of the computation tasks that offloaded to MEC server; when $\lambda = 0$ or $\lambda = 1$, it is binary offloading; when $\lambda \in (0, 1)$, it is partial offloading.

Based on Definition 1, the ratio of the computation task that executed locally is $1 - \lambda$. The transmission power of the mobile user can be adjusted from $p_{min}$ to $p_{max}$. In wireless network, the minimum transmission power $p_{min}$ should make the SINR larger than the threshold [18] (the SINR threshold relates to the hardware architecture of users). Note that when $\lambda = 0$, the computation task will be executed locally, which means $p = 0$; when $\lambda \neq 0$, the task needs to be offloaded to the MEC server and $p \neq 0$; so the feasible region of the transmission power, which takes the task offloading decision into account, is $p \in \{0, [p_{min}, p_{max}]\}$. As introduced in [18], [19], and [20], since the BS has more global information about the small cell than the mobile users, so the BS selects a spectrum band and schedules the users in its cell to different subchannels which are orthogonal. Therefore, the offloading ratio is decided by the mobile user and the communication channels of the mobile users will be allocated by the BSs in this paper.

*A. Communication Model*

In OFDMA, the mobile users in the same small cell cannot affect each other due to the orthogonality of the communication channels; only the mobile users in different small cells and use the same subchannel can affect each other. So we define the interference users set of user *n* in channel *k* as: *the set of mobile users which in different small cells with user n and use channel k to transmit data to the MEC server, denoted as* $I_n^k$. Therefore, the interference of user *n* when offloading the task to MEC server through channel *k* is: $\Gamma_n^k = \sum_{i \in I_n^k, \lambda_i > 0} p_i G_i$. Based on the Shannon-Hartley formula [24], the transmission rate of mobile user *n* can be calculated as:

$$r_n^k(\lambda, p) = \omega_k \log_2(1 + p_n G_n / (N_0 + \Gamma_n^k)) \tag{1}$$

where $\omega_k$ is the wireless channel bandwidth; $N_0$ is the power of Gaussian white noise; $G_n$ is the channel gain between mobile user $n$ and BS; $\boldsymbol{\lambda}$ is the set of offloading decisions of the mobile users and $\boldsymbol{\lambda} = \{\lambda_1, \lambda_2, \dots, \lambda_N\}$; $\boldsymbol{p}$ is the set of transmission powers of the mobile users and $\boldsymbol{p} = \{p_1, p_2, \dots, p_N\}$. From (1), we can find that the transmission rate of mobile user $n$ not only relates to its transmission power and offloading decision but also the interference users'. The transmission rate of user $n$ is affected by the interference users, and user $n$ also affects the transmission rate of the interference users. Due to the interaction between user $n$ and its interference users, the power control and the offloading decision are coupled.

B. Interference Graph

In order to quantify the inter-cell interference, we employ the interference graph defined in [18], [19], and [20]. In OFDMA, if user $i \in I_n^k$ can affect the data transmission of user $n$, then $R_i \geq |in|$, where $R_i$ is the transmission range of user $i$ and $|in|$ is the distance between user $i$ and user $n$. Based on this conclusion, two kinds of neighbor sets of user $n$ can be defined: (1) the users in $I_n^k$ and can affect the data transmission of user $n$ are defined as $I_n^{k,in}$, denoted as $I_n^{k,in} \triangleq \{i \in I_n^k, R_i \geq |in|\}$; obviously, $I_n^{k,in} \in I_n^k$; (2) the users in $I_n^k$ and can be affected by the transmission of user $n$ are defined as $I_n^{k,out}$, denoted as $I_n^{k,out} \triangleq \{i \in I_n^k, R_n \geq |in|\}$; moreover, $I_n^{k,out} \in I_n^k$. Note that the $I_n^{k,in}$ may not equal to the $I_n^{k,out}$, since the transmission powers of the mobile users are different. User $i$ can affect user $n$ does not mean that node $n$ can also affect user $i$. So the interference $\Gamma_n^k$ can be rewritten as: $\Gamma_n^k = \sum_{i \in I_n^{k,in}, \lambda_i > 0} p_i G_i$.

B. Computation Model

For mobile user $n$, there is a computation task $T_n = \{L_n, C_n\}$ which can be calculated locally or offloaded to the MEC server through BS. The $L_n$ is the length of the input data (bits) and $C_n$ is the computation workload (CPU cycles needed for one-bit data).

### B.1 Computation Model of Local Execution

When $\lambda_n = 0$, the computation task $T_n$ is executed locally, then $p_n = 0$ and $r_n(\lambda, p) = 0$; thus, the latency for calculating this task can be shown as:

$$t_n^{local} = L_n C_n / f_n \qquad (2)$$

where $f_n$ is the CPU cycles per second (i.e. the computation capability) of user $n$ with the upper bound is $f_{max}$. Different mobile users have different computation capabilities and each mobile user can adjust its computation capability from 0 to $f_{max}$. Based on [5] and [13], the energy consumption for calculating the computation task $T_n$ is:

$$e_n^{local} = \kappa_n L_n C_n f_n^2 \qquad (3)$$

In (3), $\kappa_n$ is a constant related to the hardware architecture of mobile device $n$. The dynamic power consumption is proportional to $V_c^2 f_n$, where $V_c$ is the circuit supply voltage. When the operating voltage is low, the CPU frequency is approximately linear proportional to the voltage supply [13][25][26]. Thus, the energy consumption of one CPU cycle is $\kappa_n f_n^2$. According to (2) and (3), based on the overhead defined in [5], the overhead of computing locally is:

$$O_n^{local} = \alpha_{t,n}(L_n C_n / f_n) + \alpha_{e,n} \kappa_n L_n C_n f_n^2 \qquad (4)$$

where $\alpha_{t,n}, \alpha_{e,n} \in [0,1]$ denote the weights of computational latency and energy for user $n$'s decision making, respectively. When user $n$ cares about energy consumption, the user $n$ can set $\alpha_{t,n} = 0$ and $\alpha_{e,n} = 1$; otherwise, when user $n$ is sensitive to delay, then user $n$ can set $\alpha_{t,n} = 1$ and $\alpha_{e,n} = 0$. This model can take both computational latency and energy consumption into the decision making at the same time. In practice, the proper weights can be determined by the multi-attribute utility approach in *multiple criteria decision making theory* [27].

### B.2 Computation Model of Cloud computing

When $\lambda_n > 0$, part of or all of the computation tasks are offloaded to the MEC server, where $p_n \in [p_{min}, p_{max}]$ and $r_n(\lambda, p) > 0$. Assuming that the communication channel assigned to mobile user $n$ is $k$, the latency for transmitting the input data to MEC server can be calculated as:

$$t_{n,trans}^k = L_n / r_n^k(\lambda, p) \tag{5}$$

If the transmission power of mobile user $n$ for transmitting the computation task to MEC server through channel $k$ is $p_n^k$, the energy consumption of the data transmission is:

$$e_{n,trans}^k = p_n^k t_{n,cloud}^k = p_n^k L_n / r_n^k(\lambda, p) \tag{6}$$

When the input data is offloaded to the MEC server, the MEC server calculates the computation task based on the input data. According to (2), the latency of task execution in MEC server is:

$$t_n^{cloud} = L_n C_n / f_c \tag{7}$$

where $f_c$ is the CPU cycles per second of MEC server. Similar to (3), the energy consumption in MEC server is:

$$e_n^{cloud} = \kappa_c L_n C_n f_c^2 \tag{8}$$

where $\kappa_c$ is a constant related to the hardware architecture of the MEC server.

Therefore, based on (5), (6), (7), and (8), the overhead of cloud computing can be calculated as:

$$O_n^{cloud} = \alpha_{t,n}(t_{n,trans}^k + t_n^{cloud}) + \alpha_{e,n}(e_{n,trans}^k + e_n^{cloud}) \tag{9}$$

According to the offloading decision ratio $\lambda_n$, the length of the input data that calculated locally is $(1 - \lambda_n)L_n$ and the length of the input data that offloaded to the MEC server is $\lambda_n L_n$ [9][11]. Thus, the whole computation overhead of user $n$, which takes the overheads of local computing and cloud computing into account, can be calculated as:

$$O_n = \alpha_{t,n}\left(\frac{\lambda_n L_n}{r_n^k(\lambda,p)} + \frac{\lambda_n L_n C_n}{f_c} + \frac{(1-\lambda_n)L_n C_n}{f_n}\right) + \alpha_{e,n}\left(\frac{\lambda_n p_n^k L_n}{r_n^k(\lambda,p)} + \lambda_n \kappa_c L_n C_n f_c^2 + (1-\lambda_n)\kappa_n L_n C_n f_n^2\right) \tag{10}$$

*C. Problem Statement*

For the distributed manner of the offloading decision and resource allocation in MEC system, each mobile user makes the offloading decision and allocates the wireless resource based on the local information to minimize the computation overhead. In this paper, the transmission power, the offloading decision, and the CPU capabilities of mobile users are the control variables; therefore, the problem needs to be solved in this paper is shown in **P1**:

$$\textbf{P1}: \min \sum_{n \in N} O_n(\lambda_n, p_n, f_n)$$

$$s.t. \quad 0 \leq \lambda_n \leq 1, n \in N \quad \text{(a)}$$

$$p_n \in \begin{cases} [p_{min}, p_{max}], & \lambda_n > 0, n \in N \\ \{0\}, & \lambda_n = 0, n \in N \end{cases} \quad \text{(b)} \quad (11)$$

$$0 \leq f_n \leq f_{max}, n \in N \quad \text{(c)}$$

The main issue needed to be solved is that how the mobile users choose the values of these variables in a decentralized manner to make the network overhead shown in (11) minimize. According to the conclusion in [5] and [15], the offloading decision and transmission power control between different users to make the network overhead minimum is a NP-hard problem; so it can be found easily that **P1** is also a NP-hard problem. These three variables are related and coupled. The $f_n$ and $p_n$ relate to $\lambda_n$: if $\lambda_n = 0$, then $f_n \neq 0$ and $p_n = 0$; otherwise, $f_n = 0$ and $p_n \in [p_{min}, p_{max}]$. The $\lambda_n$ determines by the values of $O_n^{cloud}$ and $O_n^{local}$ which also relates to $f_n$ and $p_n$. Moreover, the offloading decision and transmission power between different mobile users are also relevant, which can be found in (1). The centralized manners of determining the offloading decision and the transmission power has been introduced in [13], [14], and [15]. However, for the point of view of the mobile user, the centralized manner is not always effective in practice; moreover, it is not feasible for the mobile users to get the global information of the network. Thus, we introduce the game theory into the network overhead minimization.

III. MULTI-USER OFFLOADING DECISION AND RESOURCE ALLOCATION GAME

According to the **P1** that shown in (11), we define the multi-user game as: $G = (N, \{S_n\}_{n \in N}, \{O_n\}_{n \in N})$, where $S_n = \lambda_n \otimes p_n \otimes f_n$ is the strategy space of mobile user $n$; $\lambda_n$, $p_n$, and $f_n$ are the strategy spaces of offloading decision, transmission power, and CPU capability of user $n$, respectively, i.e., $\lambda_n \in \lambda_n$, $p_n \in p_n$, and $f_n \in f_n$. For each strategy of user $n$, $s_n \in S_n$ and $s_n = \{\lambda_n, p_n, f_n\}$. Moreover, $S = \{S_1, S_2, ..., S_N\}$ is the strategy space set of all the mobile users. The utility function of mobile user $n$ is $O_n(s_n, s_{-n})$, in which $s_{-n} = (s_1, s_2, ..., s_{n-1}, s_{n+1}, ..., s_N)$ is the strategies of all the others mobile users except user $n$. So the game $G$ can be stated as that for given $s_{-n}$, the mobile user $n$ would like to choose a proper strategy $s_n = \{\lambda_n, p_n, f_n\}$ to minimize its computation overhead, i.e.,

$$G: \quad \min_{s_n \in S_n} O_n(s_n, s_{-n}), \forall n \in N \tag{12}$$

According to the conclusions in [19], [20], and [28], the approach which defines the individual computation overhead as the utility of each user is selfish and cannot guarantee obtain global optimization. So, the local altruistic behaviors among neighboring users, which is motivated by local cooperation in biographical systems [29][30], is introduced into the utility function construction. In game $G$, when the strategy of user $n$ changes, the computation overheads of the users in $I_n^{k,out}$ are affected; therefore, in the new overhead function, the computation overhead of the users in $I_n^{k,out}$ are taken into account, which can be expressed as:

$$U_n(s_n, s_{-n}) = O_n(s_n, s_{-n}) + \sum_{i \in I_n^{k,out}} O_i(s_{i,s_n}, s_{-i}) \tag{13}$$

where $s_{i,s_n}$ means the strategy of user $i \in I_n^{k,out}$ when the strategy of user $n$ is $s_n$. In (13), the first term in the right hand is the computation overhead of user $n$; the second term is the aggregated computation overhead of the users in $I_n^{k,out}$. Then the new local altruistic game $G'$ is:

$$G': \quad \min_{s_n \in S_n} U_n(s_n, s_{-n}), \forall n \in N \tag{14}$$

Based on the definition of game $G'$ in (14), we define the NE of game $G'$ as follows.

**Definition 2.** A strategy profile $s^* = \{s_1^*, s_2^*, \ldots, s_N^*\}$ is a NE of game $G'$, if at the NE point $s^*$, no mobile user can reduce their computation overhead by changing the strategy unilaterally. The mathematic formula expression is: for $\forall s_n \in S_n$ and $\forall n \in N$, $U_n(s_n^*, s_{-n}^*) \leq U_n(s_n, s_{-n}^*)$ holds.

At the NE point, each user cannot find a better strategy than the current one when the other users do not change their strategies. This property is important to the distributed issues, since each mobile user minimize their own overhead based on their own interest. Based on the game theory, once the mobile users make a decision to reduce the overhead, they must take the other mobile users' strategies into account.

*A. The Existence and Uniqueness of the Nash Equilibrium*

Since the computation overhead $U_n(s_n, s_{-n})$ and the strategy space of transmission power $p_n$ are not continuous, so we introduce the concept of the potential game into the proof of the existence and uniqueness of the NE. For proving that $G'$ is a potential game, first, we define the potential game in Definition 3.

**Definition 3** [31][32]. A game is an exact potential game if there exists a potential function $\Phi$ such that for $\forall n \in N$ and $\forall s_n, s_n', s_{-n} \in S_n$, the following conditions hold:

$$U_n(s_n', s_{-n}) - U_n(s_n, s_{-n}) = \Phi(s_n', s_{-n}) - \Phi(s_n, s_{-n}) \tag{15}$$

*Remark 1:* For the exact potential game, if any mobile user changes its strategy (i.e., from $s_n$ to $s_n'$), the variation in the overhead function equals to that in the potential function. The important property of the potential game is that there always exists a NE and the asynchronous better response update process must be finite and leads to a NE [31][32].

**Theorem 1.** The game $G'$ shown in (14) is an exact potential game.

*Proof.* Based on the conclusions in [19], [20], and [28], we define the potential function as:

$$\Phi(s_n, s_{-n}) = \sum_{n \in N} O_n(s_n, s_{-n})$$

$$= O_n(s_n, s_{-n}) + \sum_{i \in I_n^{k,out}} O_i(s_{i,s_n}, s_{-i}) + \sum_{j \in N \setminus I_n^{k,out} \setminus \{n\}} O_j(s_{j,s_n}, s_{-j}) \quad (16)$$

The (16) can be divided into three parts. The first term in the right hand is the computation overhead of mobile user $n$; the second term represents the summary of the overheads of the users in $I_n^{k,out}$; the third term is the computation overhead of the rest mobile users in the network. Since there are three variables in $s_n$, so there will be $\sum_{i=1}^{3} C_3^i = 7$ different changing models of $s_n$. If we calculate the variation of the computation overhead function and the potential function based on all the changing models of $s_n$, the computation will be complex. However, the variables in $s_n$ can be divided into two different groups: 1) $\lambda_n$ and $p_n$; these two variables can not only affect the overhead of user $n$ but also the users in $I_n^{k,out}$, denoted as $par\mathrm{I}$; 2) $f_n$; when this variable changes, only the overhead of user $n$ is affected, denoted as $par\mathrm{II}$.

When $par\mathrm{I}$ changes, i.e., $\lambda_n$ or $p_n$ changes or both of these two variables change, no matter the $par\mathrm{II}$ changes or not, the overheads of user $n$ and the users in $I_n^{k,out}$ change; the computation overheads of the rest users (i.e., the users in $N \setminus I_n^{k,out} \setminus \{n\}$) do not change, because the strategy changing of user $n$ cannot affect the mobile users which are not in $I_n^{k,out}$. The deviation of the computation overhead when the user $n$'s strategy changes from $s_n$ to $s_n'$ can be calculated:

$$U_n(s_n', s_{-n}) - U_n(s_n, s_{-n}) = O_n(s_n', s_{-n}) - O_n(s_n, s_{-n})$$
$$+ \sum_{i \in I_n^{k,out}} O_i(s_{i,s_n'}, s_{-i}) - \sum_{i \in I_n^{k,out}} O_i(s_{i,s_n}, s_{-i}) \quad (17)$$

According to (16), the variation of the potential function is:

$$\Phi(s_n', s_{-n}) - \Phi(s_n, s_{-n}) = O_n(s_n, s_{-n}) - O_n(s_n', s_{-n}) + \sum_{i \in I_n^{k,out}} O_i(s_{i,s_n'}, s_{-i})$$
$$- \sum_{i \in I_n^{k,out}} O_i(s_{i,s_n}, s_{-i}) + \sum_{j \in N \setminus I_n^{k,out} \setminus \{n\}} O_j(s_{j,s_n'}, s_{-j}) - \sum_{j \in N \setminus I_n^{k,out} \setminus \{n\}} O_j(s_{j,s_n}, s_{-j}) \quad (18)$$

Since in (19), $\sum_{j \in N \setminus I_n^{k,out} \setminus \{n\}} O_j(s_{j,s_n'}, s_{-j}) - \sum_{j \in N \setminus I_n^{k,out} \setminus \{n\}} O_j(s_{j,s_n}, s_{-j}) = 0$, so $U_n(s_n', s_{-n}) - U_n(s_n, s_{-n}) = \Phi(s_n', s_{-n}) - \Phi(s_n, s_{-n})$ holds.

When the $parII$ changes and the $parI$ does not change, the computation overhead of the other mobile users except user $n$ are keep constant. So based on (17), $\sum_{i \in I_n^{k,out}} O_i(s_{i,s_n'}, s_{-i}) - \sum_{i \in I_n^{k,out}} O_i(s_{i,s_n}, s_{-i}) = 0$ and $\sum_{j \in N \setminus I_n^{k,out} \setminus \{n\}} O_j(s_{j,s_n'}, s_{-j}) - \sum_{j \in N \setminus I_n^{k,out} \setminus \{n\}} O_j(s_{j,s_n}, s_{-j}) = 0$. Thus, the conclusion in (15) holds. Therefore, Theorem 1 is proved.

**Theorem 2.** The NE of game $G'$ exists and is unique.

*Proof.* This can be proved based on the properties of the potential game shown in [31] and [32]: *if the game is an exact potential game, then it has unique pure NE strategy*. Since the game $G'$ is an exact potential game, so the NE exists and is unique. Thus, the Theorem 2 holds.

*B. The best response strategy*

For the game which the existence and uniqueness are guaranteed, the best response dynamic always converges to a NE [33][34], so the best response dynamic is applied to reach the NE of game $G'$. In the best response strategy, each mobile user calculates the best response of the variables in $s$ according to the information that gotten from BS, i.e., for given $s_{-n}$, user $n$ calculates the best response of $\lambda_n$, $p_n$, and $f_n$ based on (13).

**Corollary 1.** For $\forall p_n \in \boldsymbol{p}_n$ and $\forall f_n \in \boldsymbol{f}_n$, the best response of $\lambda_n$ will be $\lambda_n = 0$ or $\lambda_n = 1$.

*Proof.* Based on $O_n(s_n, s_{-n})$, the computation overhead function of game $G'$ can be expressed as:

$$U_n(s_n, s_{-n}) = \alpha_{t,n} \left( \frac{\lambda_n L_n}{r_n^k(\lambda, p)} + \frac{\lambda_n L_n C_n}{f_c} + \frac{(1-\lambda_n) L_n C_n}{f_n} \right) + \alpha_{e,n} \left( \frac{\lambda_n p_n^k L_n}{r_n^k(\lambda, p)} + \lambda_n \kappa_c L_n C_n f_c^2 + (1-\lambda_n) \kappa_n L_n C_n f_n^2 \right)$$

$$+ \sum_{i \in I_n^{k,out}} \left( \alpha_{t,i} \left( \frac{\lambda_i L_i}{r_i^k(\lambda, p)} + \frac{\lambda_i L_i C_i}{f_c} + \frac{(1-\lambda_i) L_i C_i}{f_i} \right) + \alpha_{e,i} \left( \frac{\lambda_i p_i^k L_i}{r_i^k(\lambda, p)} + \lambda_i \kappa_c L_i C_i f_c^2 + (1-\lambda_i) \kappa_i L_i C_i f_i^2 \right) \right) \quad (19)$$

Since when $\lambda_n = 0$, $p_n = 0$, when $\lambda_n \neq 0$, $p_n \in [p_{min}, p_{max}]$, so the $U_n(s_n, s_{-n})$ is not continuous. For solving this issues, the value of $\lambda_n$ is divided into two parts: 1) $\lambda_n \in [\varepsilon, 1]$, where $\varepsilon$ is small enough and can near to 0 arbitrarily, then $p_n \in [p_{min}, p_{max}]$; 2) $\lambda_n = 0$, then $p_n = 0$. When $\lambda_n \in [\varepsilon, 1]$ and $p_n \neq 0$, the third term in (19) has no relation with $\lambda_n$. Obviously, the $O_n(s_n, s_{-n})$ is a linear function on $\lambda_n$; so the extreme value of $O_n(s_n, s_{-n})$ will be gotten when $\lambda_n = \varepsilon$ or $\lambda_n = 1$. When $\alpha_{t,n}\left(\frac{L_n}{r_n^k(\lambda,p)} + \frac{L_n C_n}{f_c}\right) + \alpha_{e,n}\left(\frac{p_n^k L_n}{r_n^k(\lambda,p)} + \kappa_c L_n C_n f_c^2\right) > \alpha_{t,n}\frac{L_n C_n}{f_n} + \alpha_{e,n}\kappa_n L_n C_n f_n^2$, the $O_n(s_n, s_{-n})$ is an increasing function and the minimum computation overhead can be gotten when $\lambda_n = \varepsilon$. Thus, when $\varepsilon \to 0$, the $U_n(s_n, s_{-n})$ can be calculated as:

$$\lim_{\varepsilon \to 0} U_n(s_n, s_{-n}) = \alpha_{t,n}\frac{L_n C_n}{f_n} + \alpha_{e,n}\kappa_n L_n C_n f_n^2 + \sum_{i \in I_n^{k,out}} \left(\alpha_{t,i}\left(\frac{\lambda_i L_i}{r_{i,p_n \neq 0}^k(\lambda,p)} + \frac{\lambda_i L_i C_i}{f_c} + \frac{(1-\lambda_i)L_i C_i}{f_i}\right)\right.$$

$$\left. + \alpha_{e,i}\left(\frac{\lambda_i p_i^k L_i}{r_{i,p_n \neq 0}^k(\lambda,p)} + \lambda_i \kappa_c L_i C_i f_c^2 + (1-\lambda_i)\kappa_i L_i C_i f_i^2\right)\right) \quad (20)$$

When $\alpha_{t,n}\left(\frac{L_n}{r_n^k(\lambda,p)} + \frac{L_n C_n}{f_c}\right) + \alpha_{e,n}\left(\frac{p_n^k L_n}{r_n^k(\lambda,p)} + \kappa_c L_n C_n f_c^2\right) < \alpha_{t,n}\frac{L_n C_n}{f_n} + \alpha_{e,n}\kappa_n L_n C_n f_n^2$, the $O_n(s_n, s_{-n})$ is a decreasing function; so the minimum overhead will be gotten when $\lambda_n = 1$, which is:

$$U_n(s_n, s_{-n})|_{\lambda_n=1} = \alpha_{t,n}\left(\frac{L_n}{r_n^k(\lambda,p)} + \frac{L_n C_n}{f_c}\right) + \alpha_{e,n}\left(\frac{p_n^k L_n}{r_n^k(\lambda,p)} + \kappa_c L_n C_n f_c^2\right)$$

$$+ \sum_{i \in I_n^{k,out}} \left(\alpha_{t,i}\left(\frac{\lambda_i L_i}{r_{i,p_n \neq 0}^k(\lambda,p)} + \frac{\lambda_i L_i C_i}{f_c} + \frac{(1-\lambda_i)L_i C_i}{f_i}\right)\right.$$

$$\left. + \alpha_{e,i}\left(\frac{\lambda_i p_i^k L_i}{r_{i,p_n \neq 0}^k(\lambda,p)} + \lambda_i \kappa_c L_i C_i f_c^2 + (1-\lambda_i)\kappa_i L_i C_i f_i^2\right)\right) \quad (21)$$

When $\lambda_n = 0$, the computation overhead shown in (19) can be calculated as:

$$U_n(s_n, s_{-n})|_{\lambda_n=0} = \alpha_{t,n}\frac{L_n C_n}{f_n} + \alpha_{e,n}\kappa_n L_n C_n f_n^2 + \sum_{i \in I_n^{k,out}} \left(\alpha_{t,i}\left(\frac{\lambda_i L_i}{r_{i,p_n=0}^k(\lambda,p)} + \frac{\lambda_i L_i C_i}{f_c} + \frac{(1-\lambda_i)L_i C_i}{f_i}\right)\right.$$

$$\left. + \alpha_{e,i}\left(\frac{\lambda_i p_i^k L_i}{r_{i,p_n=0}^k(\lambda,p)} + \lambda_i \kappa_c L_i C_i f_c^2 + (1-\lambda_i)\kappa_i L_i C_i f_i^2\right)\right) \quad (22)$$

According to (1), we have $r_{i,p_n \neq 0}^k(\lambda, p) = \omega_k \log_2\left(1 + \frac{p_i^k G_i}{N_0 + \sum_{j \in I_i^k} p_j^k G_j + p_n^k G_n}\right)$ and $r_{i,p_n=0}^k(\lambda, p) = \omega_k \log_2\left(1 + \frac{p_i^k G_i}{N_0 + \sum_{j \in I_i^k} p_j^k G_j}\right)$, so $r_{i,p_n \neq 0}^k(\lambda, p) < r_{i,p_n=0}^k(\lambda, p)$; therefore, based on (20) and (22), $\lim_{\varepsilon \to 0} U_n(s_n, s_{-n}) > U_n(s_n, s_{-n})|_{\lambda_n=0}$. So in this case, the best response of $\lambda_n$ is $\lambda_n = 0$.

When $\alpha_{t,n}\left(\frac{L_n}{r_n^k(\boldsymbol{\lambda},\boldsymbol{p})}+\frac{L_nC_n}{f_c}\right)+\alpha_{e,n}\left(\frac{p_n^kL_n}{r_n^k(\boldsymbol{\lambda},\boldsymbol{p})}+\kappa_cL_nC_nf_c^2\right)<\alpha_{t,n}\frac{L_nC_n}{f_n}+\alpha_{e,n}\kappa_nL_nC_nf_n^2$, the overhead function shown in (19) is a decreasing function. On one hand, the first two terms in (21) are smaller than that in (22); on the other hand, the third term in (21) is larger than that in (22) due to $r_{i,p_n\neq0}^k(\boldsymbol{\lambda},\boldsymbol{p})<r_{i,p_n=0}^k(\boldsymbol{\lambda},\boldsymbol{p})$. This means that $U_n(s_n,s_{-n})|_{\lambda_n=1}$ may larger or smaller than $U_n(s_n,s_{-n})|_{\lambda_n=0}$. Therefore, in this case, the best response of $\lambda_n$ is $\lambda_n=1$ or $\lambda_n=0$ according to the values of $U_n(s_n,s_{-n})|_{\lambda_n=1}$ and $U_n(s_n,s_{-n})|_{\lambda_n=0}$. Thus, the Corollary 1 holds.

*Remark 2:* In the proof of Corollary 1, when $\alpha_{t,n}\left(\frac{L_n}{r_n^k(\boldsymbol{\lambda},\boldsymbol{p})}+\frac{L_nC_n}{f_c}\right)+\alpha_{e,n}\left(\frac{p_n^kL_n}{r_n^k(\boldsymbol{\lambda},\boldsymbol{p})}+\kappa_cL_nC_nf_c^2\right)>\alpha_{t,n}\frac{L_nC_n}{f_n}+\alpha_{e,n}\kappa_nL_nC_nf_n^2$, the value of $\lambda_n$ is $\lambda_n=0$. This is easy to be understood since in this case, the task is calculated locally costs less resource than that in MEC server. However, when $\alpha_{t,n}\left(\frac{L_n}{r_n^k(\boldsymbol{\lambda},\boldsymbol{p})}+\frac{L_nC_n}{f_c}\right)+\alpha_{e,n}\left(\frac{p_n^kL_n}{r_n^k(\boldsymbol{\lambda},\boldsymbol{p})}+\kappa_cL_nC_nf_c^2\right)<\alpha_{t,n}\frac{L_nC_n}{f_n}+\alpha_{e,n}\kappa_nL_nC_nf_n^2$, the best response of $\lambda_n$ is $\lambda_n=1$ or $\lambda_n=0$; this means that even the user $n$'s computation overhead of cloud computing is smaller than that of local computing, considering the overheads of the users in $I_n^{k,out}$, the user $n$ may not offloads its task to MEC server.

**Corollary 2.** The best response of the CPU capability $f_n$ is $f_n=\left(\alpha_{t,n}/2\alpha_{e,n}\kappa_n\right)^{1/3}$ or $f_n=0$.

*Proof.* When $\lambda_n\neq0$, $f_n=0$; only when $\lambda_n=0$, $f_n\in(0,f_{max}]$. Thus, when $\lambda_n=1$, the best response of $f_n$ is $f_n=0$. When $\lambda_n=0$, the computation overhead function is shown in (22). The first derivative of (22) on $f_n$ is: $U_n'(s_n,s_{-n})|_{f_n}=2\alpha_{e,n}\kappa_nL_nC_nf_n-\alpha_{t,n}\frac{L_nC_n}{f_n^2}$. When $U_n'(s_n,s_{-n})|_{f_n}=0$, $f_n=\left(\alpha_{t,n}/2\alpha_{e,n}\kappa_n\right)^{1/3}$; moreover, since $U_n''(s_n,s_{-n})|_{f_n}=2\alpha_{e,n}\kappa_nL_nC_n+2\alpha_{t,n}\frac{L_nC_n}{f_n^3}>0$, so when $f_n=\left(\alpha_{t,n}/2\alpha_{e,n}\kappa_n\right)^{1/3}$, the $U_n(s_n,s_{-n})$ can get the minimum value. Thus the best response of the CPU capability $f_n$ is $f_n=\left(\alpha_{t,n}/2\alpha_{e,n}\kappa_n\right)^{1/3}$ or $f_n=0$.

**Corollary 3.** For $\forall \lambda_n \in \{0,1\}$, the best response of $p_n$ exists and $p_n = 0$ or $p_n = \bar{p}_n$, where $\bar{p}_n = \arg \min_{n \in N}\{U_n(s_{n|p_n=p_{min}}, s_{-n}), U_n(s_{n|p_n=p'_n}, s_{-n}), U_n(s_{n|p_n=p_{max}}, s_{-n})\}$.

*Proof.* Since the best response of offloading decision is $\lambda_n \in \{0,1\}$ and when $\lambda_n = 0$, $p_n = 0$, so the best response of $p_n$ when $\lambda_n = 0$ is $p_n = 0$. When $\lambda_n = 1$, the task is offloaded to the MEC server. Since the delay and the energy consumption when the task is executed in MEC server have no relation with the transmission power, so we only consider the latency and the energy consumption caused by the data transmission. According to (19), the transmission overhead can be calculated as $U_n(s_n, s_{-n})_{trans} = \alpha_{t,n}\frac{L_n}{r_n^k(\lambda,p)} + \alpha_{e,n}\frac{p_n^k L_n}{r_n^k(\lambda,p)} + \sum_{i \in I_n^{k,out}} U_i(s_i, s_{-i})$; considering (1), the overhead of transmission can be written as:

$$U_n(s_n, s_{-n})_{trans} = a_n \frac{\alpha_{t,n} + \alpha_{e,n}p_n^k}{\ln(1+(p_n^k G_n/\Gamma_n^k))} + \sum_{i \in I_n^{k,out}} \frac{b_i}{\ln\left(1+\left(p_i^k G_i/N_0+p_n^k G_n+\sum_{j \in I_i^{k,in}_{\{n\}}} p_j^k G_j\right)\right)} \quad (23)$$

where $a_n = L_n \ln 2/\omega_k$, $b_i = \lambda_i L_n(\alpha_{t,i} + \alpha_{e,i}p_i^k)\ln 2/\omega_k$. The extreme value of (23) can be gotten when $U'_n(s_n, s_{-n})_{trans}|_{p_n^k} = 0$. Since $U'_n(s_n, s_{-n})_{trans}|_{p_n^k} = 0$ is a transcendental equation, so the analytical solution of $U'_n(s_n, s_{-n})_{trans}|_{p_n^k} = 0$ does not exist. The numerical solution of $U'_n(s_n, s_{-n})_{trans}|_{p_n^k} = 0$ can be calculated by Newton Method. Let $p'_n$ be the solution of $U'_n(s_n, s_{-n})_{trans}|_{p_n^k} = 0$; since the solution of $U'_n(s_n, s_{-n})_{trans}|_{p_n^k} = 0$ may not single, so we define the $p'_n$ as: $p'_n = \arg\{U'_n(s_n, s_{-n})_{trans}|_{p_n^k} = 0\}$. Therefore, if $p'_n = \emptyset$, the $U_n(s_n, s_{-n})_{trans}$ is a monotone function with $p_n \in [p_{min}, p_{max}]$, then the best response of $p_n$ will be $p_n = p_{min}$ or $p_n = p_{max}$. If $p'_n \neq \emptyset$, the extreme value of (23) will be gotten at $p'_n$; therefore, if $p'_n \in [p_{min}, p_{max}]$, it means that the extreme value of (23) exists in the feasible region $[p_{min}, p_{max}]$; if $p'_n \notin [p_{min}, p_{max}]$, the extreme value of (24) can be gotten when $p'_n = p_{min}$ or $p'_n = p_{max}$. Thus, the best response of the transmission power $p_n$ can be calculated as

$\bar{p}_n = \arg \min_{n \in N}\{U_n(s_n|_{p_n=p_{min}}, s_{-n}), U_n(s_n|_{p_n=p'_n}, s_{-n}), U_n(s_n|_{p_n=p_{max}}, s_{-n})\}$. Therefore, the best response of $p_n$ is $p_n = 0$ or $p_n = \bar{p}_n$.

**Corollary 4.** The best response of $\lambda_n$ is decided by the interference from the mobile users in $I_n^{k,out}$, $I_n^{k,in}$, and $I_i^{k,in}$, where $I_i^{k,in}$ is the set of the interference users which can affect the data transmission of user $i \in I_n^{k,out}$.

*Proof.* Since the best response of $\lambda_n$ is $\lambda_n = 0$ or $\lambda_n = 1$, so the best response of $U_n(s_n, s_{-n})$ can be expressed as:

$$U_n^*(s_n, s_{-n}) = \begin{cases} U_1^* = U_n^*(s_n|_{p_n=\bar{p}_n}, s_{-n})_{trans} + U_n^*(s_n|_{f_n=0}, s_{-n})_{cloud} & , \lambda_n = 1 \\ U_0^* = U_n^*(s_n|_{p_n=0}, s_{-n})_{trans} + U_n^*\left(s_n|_{f_n=(\alpha_{t,n}/2\alpha_{e,n}\kappa_n)^{1/3}}, s_{-n}\right)_{local} & , \lambda_n = 0 \end{cases} \quad (24)$$

where $U_n^*(s_n|_{p_n=\bar{p}_n}, s_{-n})_{trans}$ and $U_n^*(s_n|_{p_n=0}, s_{-n})_{trans}$ are the best response of $U_n(s_n, s_{-n})_{trans}$, which can be gotten when $p_n = \bar{p}_n$ and $p_n = 0$ in (23); $U_n^*(s_n|_{f_n=0}, s_{-n})_{cloud} = \alpha_{t,n} L_n C_n / f_c + \alpha_{e,n}\kappa_c L_n C_n f_c^2$ is the best response of the computation overhead in MEC server; $U_n^*\left(s_n|_{f_n=\left(\frac{\alpha_{t,n}}{2\alpha_{e,n}\kappa_n}\right)^{1/3}}, s_{-n}\right)_{local}$ is the best response of the computation overhead in mobile user $n$.

For the best response strategy, if $U_1^* < U_0^*$, then $\lambda_n = 1$; otherwise, if $U_1^* > U_0^*$, then $\lambda_n = 0$. From (24), we can conclude that the values of $U_1^*$ and $U_0^*$ relate to the best response of $\lambda_n, p_n, f_n$, and the interference from the interference users (including the users in $I_n^{k,out}$, $I_n^{k,in}$, and $I_i^{k,in}$); moreover, the value of $(\alpha_{t,n}/2\alpha_{e,n}\kappa_n)^{1/3}$ is constant and the best response of $p_n$ is determined by the interference from the users in $I_n^{k,out}$ and $I_n^{k,in}$. So we can conclude that $U_1^*$ and $U_0^*$ will be decided by the interference from the interference users. This means that the best response of $\lambda_n$ is also decided by the interference from the interference users. Thus, the Corollary 4 holds.

*Remark 3:* The Corollary 4 is easy to be understood since for the mobile user $n$, if the interference from the interference users is high, then calculating the task locally is beneficial;

otherwise, offloading the task to the MEC server is the best choice. The Corollary 4 does not mean that $\lambda_n$ and $p_n$ are independent, since both $\lambda_n$ and $p_n$ are affected by the interference caused by the data transmission of the interference users. The Corollary 4 indicates that minimizing the interference is one possible approach to improve the performance of MEC.

IV. DISTRIBUTED MULTI-USER POWER CONTROL AND OFFLOADING DECISION ALGORITHM

For enabling the mobile users to achieve a mutually satisfactory decision making for power control, offloading decision, and CPU capability adjustment, the game theory is applied. Since we have proved the existence and the uniqueness of the NE, so the main idea of this algorithm is to let the mobile users improve their strategies in each time slot and reach the NE at the end. Thus, the slotted time structure and best response dynamic are used in this algorithm.

*A. Algorithm Design*

In each time slot, before calculating the best responses of $\lambda_n$, $p_n$, and $f_n$, the BS collects the necessary information[1] from the interference users by the cooperation between different BSs [5][18][19][20]. Considering the fact that the interference users can affect the data transmission of user *n* only when they offload the computation tasks to the MEC servers, so it is beneficial if the information collection and exchanging are done by the BSs (i.e., the base stations that these interference users belong to); moreover, the resource in BS is also much richer than that in mobile user. At the beginning of each time slot, the BS *l* that user *n* belongs to collects the information of the interference users from the neighbor BSs. When the necessary information has gotten by the BS, it will send the information to user *n*. Then user *n* calculates the best responses of $\lambda_n$, $p_n$, and $f_n$, and updates its strategy based on these best responses. First, user *n* calculates the values of $\bar{p}_n$ and $\left(\alpha_{t,n}/2\alpha_{e,n}\kappa_n\right)^{1/3}$ according to Corollary 2 and Corollary 3; when these two values are gotten, the best response of $U_1^*$ and $U_0^*$ can be calculated. The offloading decision $\lambda_n$ is decided based on the rules as follows: *If $U_1^* < U_0^*$, then $\lambda_n = 1$; otherwise, if $U_1^* > U_0^*$, $\lambda_n = 0$.* Once the offloading decision is determined, the best response of the transmission power $p_n$ and the CPU capability $f_n$ can be decided based on: *If $\lambda_n = 0$, then $p_n = 0$ and $f_n = (\alpha_{t,n}/2\alpha_{e,n}\kappa_n)^{1/3}$; otherwise, if $\lambda_n = 1$, then $p_n = \bar{p}_n$ and $f_n = 0$.* This process will be executed repeatedly until the NE is reached.

---

[1] The necessary information of the interference users include all the necessary parameters to calculate $U_n(s_n, s_{-n})$ shown in (19), which includes the offloading decision, the transmission power, the size of the task data of the interference users.

---

**Algorithm 1**: Joint offloading decision and resource allocation

1. **initialization:**
2. each mobile user $n$ chooses the offloading decision, the transmission power, and CPU capability as: $\lambda_n = 1$, $p_n = p_{max}$, and $f_n = 0$;
3. **end**
4. **repeat** for each user and each decision slot in parallel:
5. send the pilot signal on the chosen communication channel to the base station;
6. receive the necessary information of the interference users of other small cells;
7. calculate the best response of the transmission power $\bar{p}_n$ and the CPU capability $(\alpha_{t,n}/2\alpha_{e,n}\kappa_n)^{1/3}$;
8. calculate the $U_1^*$ and $U_0^*$ based on $\bar{p}_n$ and $(\alpha_{t,n}/2\alpha_{e,n}\kappa_n)^{1/3}$;
9. **if** $U_1^* > U_0^*$
10.     $\lambda_n = 0$;
11. **else if** $U_1^* < U_0^*$
12.     $\lambda_n = 1$;
13. **end if**
14. **if** $\lambda_n = 0$
15.     $p_n = 0$ and $f_n = (\alpha_{t,n}/2\alpha_{e,n}\kappa_n)^{1/3}$;
16. **else if** $\lambda_n = 1$
17.     $p_n = \bar{p}_n$ and $f_n = 0$;
18. **end if**
19. **repeat until** NE is meet.

---

*B. Convergence Analysis and Computation Complexity*

In [31] and [32], the authors have proved that for any potential game, the best response dynamics always converge to a pure Nash Equilibrium. Since the game $G'$ is an exact potential game, so the algorithm proposed in this paper is convergent.

**Corollary 5.** Finding the NE of game $G'$ by the best response approach is PLS complete.

*Proof.* In [31] and [32], the authors have proved that for the potential game which applies the best response approach in the process of reaching NE, if the best response can be computed in polynomial time, then this problem is PLS (Polynomial Local Search) complete. In game $G'$, there are three best responses need to be calculated in each time slot, which are $\lambda_n$, $p_n$, and $f_n$. The computation complexity for calculating the best response of $f_n$ is $O(1)$, since the best response of $f_n$ is constant. For calculating the best response of $p_n$, the Newton Method is applied. In [35], the authors have proved that for $f(x)$, the computation complexity of the Newton Method is $O[\log(n)F(n)]$, where $F(n)$ is the computation cost of $f(x)/f'(x)$. Moreover, in game $G'$, $f(x) = U_n(s_n, s_{-n})$. Since $U_n(s_n, s_{-n})$ and $U'_n(s_n, s_{-n})$ are all polynomial, so the $U_n(s_n, s_{-n})/U'_n(s_n, s_{-n})$ is also polynomial. This means that the computation of $F(n)$ can be completed in polynomial time. Thus, the calculation of the best response of $p_n$ can be finished in polynomial time. Based on Corollary 4, the best response of $\lambda_n$ relates to the calculation of $\bar{p}_n$ and $(\alpha_{t,n}/2\alpha_{e,n}\kappa_n)^{1/3}$; therefore, the computation of $\lambda_n$ also can be completed in polynomial time. Thus, the Corollary 5 holds.

*Remark 4:* For game $G'$, in each time slot, the computation complexity relates to the best response calculation of $\lambda_n$, $p_n$, and $f_n$. As shown in Corollary 5, the complexity of the best response computation of $f_n$ is $O(1)$, which is much simpler than that of $p_n$ and $\lambda_n$ and can be ignored. Moreover, when the best response of $p_n$ and $f_n$ are gotten, the complexity of the best response calculation of $\lambda_n$ is also simple, which is shown in (24). Therefore, the main computation complexity of this algorithm is caused by the calculation of the best response of $p_n$, i.e., the Newton Method. So the computation complexity of game $G'$ is $O[\log(n)F(n)]$ in one time slot, where $F(n)$ is the computation cost of $U_n(s_n, s_{-n})/U'_n(s_n, s_{-n})$. According to (19), the computation cost of $F(n)$ is $O[n\log^2(n)]$. Therefore, the computation complexity of this algorithm in one time slot is $O[n \log^3(n)]$. Assuming that the algorithm needs $C$ rounds iteration for reaching NE, then the computation complexity of the proposed algorithm is [5]: $O[Cn \log^3(n)]$. This demonstrates that the algorithm can be completed in polynomial time.

*C. Theoretical Analysis of the proposed algorithm*

In this section, we learn the price of anarchy (PoA) of the computation overhead of the whole network, i.e., $\sum_{n \in N, s_n \in s} U_n$. Based on the conclusion in [36], the PoA is defined as:

$$PoA = \frac{\sum_{n \in N, s_n \in s} U_n(\tilde{p})}{\sum_{n \in N, s_n \in s} \overline{U}_n(p^*)} \quad (25)$$

where $\tilde{s}$ is the NE of game $G'$, $s^*$ is the centralized optimal solution for all the mobile users, $U_n$ is the overhead of user $n$ by the game theory based algorithm; $\overline{U}_n$ is the overhead by the centralized algorithm. For the MEC, the smaller PoA, the better performance of is [5].

**Corollary 6.** For game $G'$, the PoA of the network computation overhead satisfies that:

$$1 \leq PoA \leq \frac{\sum_{n=1, \lambda_n=1}^{N} \left( U_{n,c}^{max} + \sum_{i \in I_n^{k,out}} U_i \right) + \sum_{n=1, \lambda_n=0}^{N} \left( U_n^{max} + \sum_{i \in I_n^{k,out}} U_i \right)}{\sum_{n=1, \lambda_n=1}^{N} \left( \overline{U}_{n,c}^{min} + \sum_{i \in I_n^{k,out}} U_i \right) + \sum_{n=1, \lambda_n=0}^{N} \left( \overline{U}_n^{min} + \sum_{i \in I_n^{k,out}} U_i \right)} \quad (26)$$

where $U_{n,c}^{max} = \frac{(\alpha_t + \alpha_e p_n) s_n}{w_k \log_2 \left( 1 + \frac{p_n G_n}{N_0 + \sum_{j \in I_n^{k,in}, \lambda_j=1} p'_{max} G_j} \right)} + \alpha_t t_{n,c} + \alpha_e e_{n,c}$, $U_n^{max} = \frac{\alpha_t c_n}{f_n^{max}} + \alpha_e \kappa_n c_n (f_n^{max})^2$,

$\overline{U}_{n,c}^{max} = \frac{(\alpha_t + \alpha_e p_n) s_n}{w_k \log_2 \left( 1 + \frac{p_n G_n}{N_0} \right)} + \alpha_t t_{n,c} + \alpha_e e_{n,c}$, $\overline{U}_n^{max} = \frac{\alpha_t c_n}{f_n^*} + \alpha_e \kappa_n c_n (f_n^*)^2$, and $p'_{max} \triangleq \max\{p_j, j \in I_n^{k,in}\}$. The $p'_{max}$ is the maximum transmission power of all the mobile users in $I_n^{k,in}$.

*Proof.* Assuming that $\tilde{s}$ is the NE of game $G'$ and $s^*$ is the centralized optimal solution for all the mobile users. According to the conclusion in [5] and [31], since the performance of the centralized optimal algorithm is better than that of game theory based algorithm, so $PoA \geq 1$.

For the game theory based algorithm, when $\lambda_n = 1$, the transmission rate of user $n$ satisfies that:

$$r_n(\tilde{s}) = w_k \log_2 \left( 1 + \frac{p_n G_n}{N_0 + \sum_{j \in I_n^{k,in}, \lambda_j=1} p_j G_j} \right) \geq w_k \log_2 \left( 1 + \frac{p_n G_n}{N_0 + \sum_{j \in I_n^{k,in}, \lambda_j=1} p'_{max} G_j} \right) \quad (27)$$

where $p'_{max} \triangleq \max\{p_j, j \in I_n^{k,in}\}$. Based on (9) and (27), the overhead of user $n$ satisfies that:

$$U_{n,c} = \frac{(\alpha_t + \alpha_e p_n)s_n}{w_k \log_2\left(1 + \frac{p_n G_n}{N_0 + \sum_{j \in I_n^{k,in}, \lambda_j = 1} p_j G_j}\right)} + \alpha_t t_{n,c} + \alpha_e e_{n,c} \leq \frac{(\alpha_t + \alpha_e p_n)s_n}{w_k \log_2\left(1 + \frac{p_n G_n}{N_0 + \sum_{j \in I_n^{k,in}, \lambda_j = 1} p'_{max} G_j}\right)}$$
$$+ \alpha_t t_{n,c} + \alpha_e e_{n,c} = U_{n,c}^{max} \quad (28)$$

When $\lambda_n = 0$, the overhead of user $n$ can be calculated as:

$$U_n = \frac{\alpha_t c_n}{f_n^{max}} + \alpha_e \kappa_n c_n (f_n^{max})^2 \leq \frac{\alpha_t c_n}{f_n^{max}} + \alpha_e \kappa_n c_n (f_n^{max})^2 = U_n^{max} \quad (29)$$

For the centralized optimal algorithm, when $\lambda_n = 1$, the transmission rate satisfied that:

$$r_n(\boldsymbol{s}^*) = w_k \log_2\left(1 + \frac{p_n G_n}{N_0 + \sum_{j \in I_n^{k,in}, \lambda_j = 1} p_j G_j}\right) \leq w_k \log_2\left(1 + \frac{p_n G_n}{N_0}\right) \quad (30)$$

Therefore, when $\lambda_n = 1$, we have:

$$U_{n,c} = \frac{(\alpha_t + \alpha_e p_n)s_n}{w_k \log_2\left(1 + \frac{p_n G_n}{N_0 + \sum_{j \in N\setminus\{n\}, \lambda_j = 1} p_j G_j}\right)} + \alpha_t t_{n,c} + \alpha_e e_{n,c} \geq \frac{(\alpha_t + \alpha_e p_n)s_n}{w_k \log_2\left(1 + \frac{p_n G_n}{N_0}\right)} + \alpha_t t_{n,c} + \alpha_e e_{n,c} = \bar{U}_{n,c}^{min} \quad (31)$$

Similar to the game theory based algorithm, when $\lambda_n = 0$, the computation overhead is:

$$U_n = \frac{\alpha_t c_n}{f_n} + \alpha_e \kappa_n c_n f_n^2 \geq \frac{\alpha_t c_n}{f_n^{min}} + \alpha_e \kappa_n c_n (f_n^{min})^2 = \bar{U}_n^{min} \quad (32)$$

Thus, according to (28), (29), (31), and (32), the Corollary 6 holds.

Note that the PoA reflects the worst cause of the game theory based algorithm on the centralized optimal algorithm; therefore, based on Corollary 6, we can conclude that with the reducing of the interference from the interference users, the PoA decreases. This demonstrates that controlling the network interference is effective on improving the performance of the MEC system, which is consistent with the conclusion in Corollary 4.

**Corollary 7.** The NE of game $G'$ is a global optimal solution for the optimal issue shown in (11).

*Proof.* Assuming that $\boldsymbol{s} = \{s_1, s_2, \ldots, s_N\}$ is a NE of game $G'$ and $\boldsymbol{s}' = \{s'_1, s'_2, \ldots, s'_N\}$ is a global optimal solution of (11). Since $\boldsymbol{s}' = \{s'_1, s'_2, \ldots, s'_N\}$ is a global optimal solution, so we have $\sum_{n \in N} O_n(s_n, s_{-n}) \geq \sum_{n \in N} O_n(s'_n, s'_{-n})$. Moreover, $G'$ is an exact potential game with potential function is $\Phi(s_n, s_{-n}) = \sum_{n \in N} O_n(s_n, s_{-n})$, so $\Phi(s_n, s_{-n}) \leq \Phi(s'_n, s'_{-n})$ holds when $s$ is a NE and $s'$ is not, i.e., $\sum_{n \in N} O_n(s_n, s_{-n}) \leq \sum_{n \in N} O_n(s'_n, s'_{-n})$. This conclusion is contradictory with the assumption, so the Corollary 7 holds.

**Corollary 8.** The NE of game $G'$ is Pareto Efficiency.

*Proof.* Based on [37], the Pareto efficiency is defined as: if $x = \{x_1, x_2, ..., x_N\}$, where $x_i \in \mathbb{R}$ and $i \in N$, is Pareto optimal solution and $U_i$ is the utility of $i$, then there is no other feasible solution $x' = \{x'_1, x'_2, ..., x'_N\}$ such that $U_i(x'_i) \geq U_i(x_i)$ for all the users with $U_i(x'_i) > U_i(x_i)$ for some users. In game $G'$, assuming that $s = \{s_1, s_2, ..., s_N\}$ is a NE; so for $\forall n \in N$, there is no solution $s' = \{s'_1, s'_2, ..., s'_N\}$ which can make the $U_n(s_n, s_{-n}) \geq U_n(s'_n, s_{-n})$ hold; moreover, the solution $s' = \{s'_1, s'_2, ..., s'_N\}$ also cannot make the $U_n(s_n, s_{-n}) > U_n(s'_n, s_{-n})$ holds for all the users. According to Corollary 7, the NE of game $G'$ is also a global optimal solution of (11). So, on one hand, the NE of game $G'$ is the optimal solution of the global function shown in (11), which means that $U_i(x'_i) > U_i(x_i)$ for some users holds; on the other hand, no users can reduce their computation overhead without increasing other users overhead, which means that $U_i(x'_i) \geq U_i(x_i)$ for all the users holds. Thus, the NE of game $G'$ is Pareto efficiency.

## V. NUMERICAL RESULT

In this simulation, there are 5 BSs and the coverage range of each BS is 50m [5][38]; the mobile users are deployed in the coverage area of BS with the numbers vary from 20 to 50. The bandwidth of the wireless channel is 5MHz. The transmission power of user changes from $p_{min}$ to $p_{max}$; the $p_{min}$ can be calculated according to the SINR threshold and the measured interference in Section IV; the $p_{max}$ is set to 150mW. The noise is -100dBm [26]. The channel gain $G_n = d_{n,l}^{-\gamma}$ [26], where $d_{n,l}$ is the distance between the mobile user $n$ and BS $l$; $\gamma$ is the path loss factor which is set to 4. Similar to [5], in this simulation, $L_n = 5000$Kb and $C_n = 1000$Megacycles. The

CPU computation capability $f_c$ is 10GHz. The decision weights $\alpha_t, \alpha_e \in [0,1]$ and $\alpha_t + \alpha_e = 1$, so we set $\alpha_t \in \{1, 0.5, 0\}$ [5]. For each mobile user, $\kappa_n = 10^{-27}$ and $f_{n,max} = 1\text{GHz}$ [13].

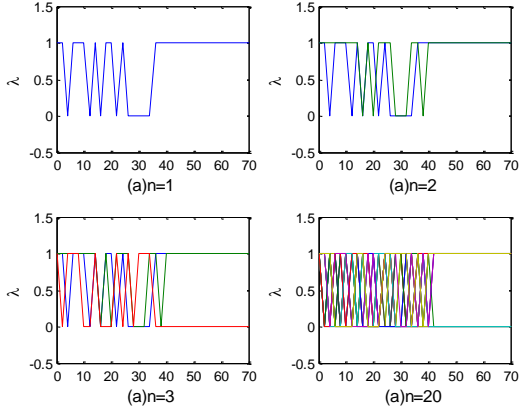

Fig.2. Convergence of offloading decision

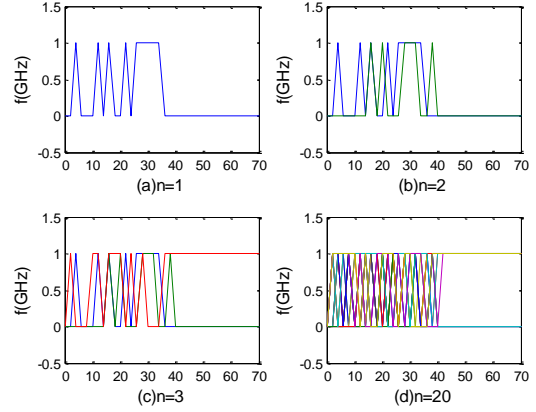

Fig.3. Convergence of the CPU capability

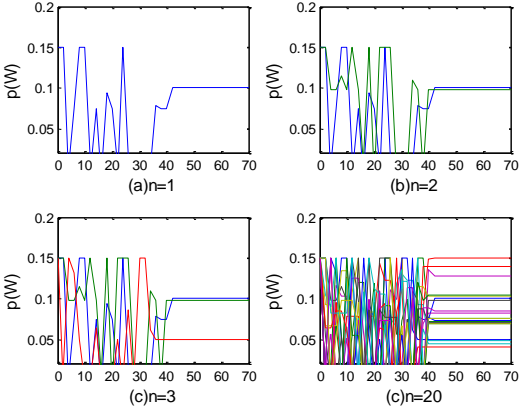

Fig.4. Convergence of the transmission power

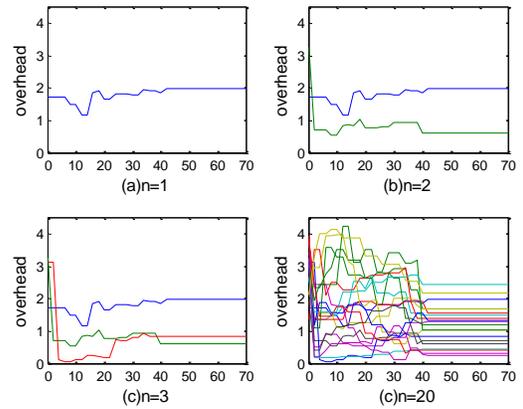

Fig.5. Convergence of the user's computation overhead

Fig.2 to Fig.5 illustrates the convergence of the algorithm proposed in this paper. Fig.2 shows the convergence of the offloading decisions. The same as that proved in Corollary 1, the value of λ converges to 0 or 1 with the increasing of the iteration times. As shown in Fig.3, the value of CPU capability is also binary, i.e. $f = 0$ or $f = f_{max}$. Moreover, the values of the CPU capability and the offloading decision are contrary, which can be found in Fig.2 and Fig.3. This is consistent with the theoretical analysis in Section III. The transmission power of mobile user is shown in Fig.4, which is convergence and smaller than $p_{max}$. Different with the CPU capability, when the offloading decision is 1, the transmission power is larger than 0; otherwise, the transmission power equals to 0. Fig.5 demonstrates that the computation overhead of each mobile user is convergent, too.

Fig.6 demonstrates the variation of the PoA with the increasing of the interference. In this simulation, we use $1/SINR$ to represent the effection of the interference, where $SINR = \sum_{n \in N} SINR_n$ and $SINR_n$ is the SINR of user $n$. So the larger $1/SINR$, the larger interference is. With the increasing of the interference, the PoA increases. This can be concluded based on Corollary 6. When the value of $1/SINR$ is fixed, such as 0.7, the PoA decreases with the increasing of the network density. This is because when the value of $1/SINR$ is fixed, the more users in the network, the smaller interference of each user; so the PoA is small when the network density is large. Fig.7 illustrates the number of users which offload their computation tasks to the MEC server under different network densities and computation loads. Since the algorithm is convergence, so the number of users which offload task to MEC server becomes constant when the iteration times is large enough (the more users in the network, the more iteration times needed). Before reaching the NE point, with the increasing of the iteration time, the resource in mobile users is consumed and more and more mobile user chooses to offload the computation task to the MEC server. This is because the MEC server has much better computation capability and resource than the mobile users. The number of users which offload tasks to the MEC server increases when the network density increases. However, this increasing becomes slowly when the network density is large; because, the inter-cell interference is serious when the network density is large, so the ratio of users which offload their tasks to MEC server decreases. Moreover, if the length of input data increases, the number of users which offload the tasks to MEC server under the same network density increases, too. This is because when the length of the data increases, the computation overhead of the local execution increases, which can be found in (24); therefore, more users will offload their computation task to the MEC server for saving energy and reducing latency.

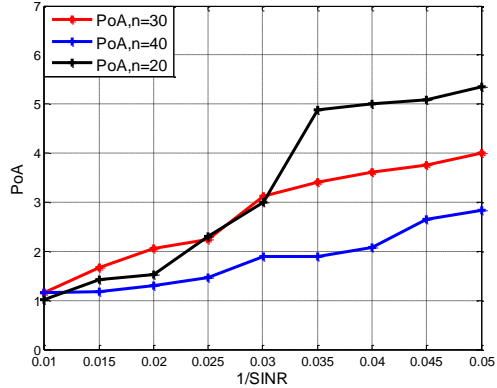

Fig.6. The PoA under different network conditions

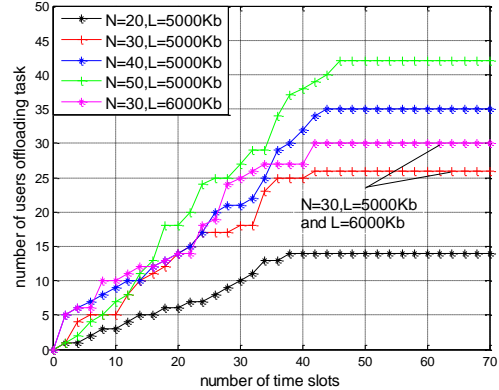

Fig.7. The number of beneficial users under different network conditions

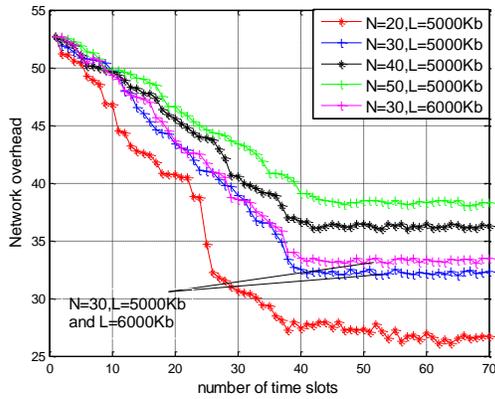

Fig.8. The network overhead under different network conditions

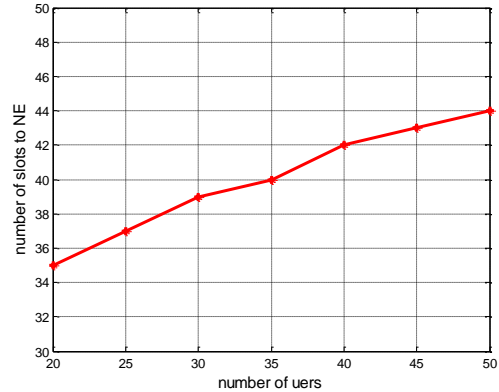

Fig.9. The convergence rate under different network density

In Fig.8, the network overheads under different network densities, different computation loads, and different iteration times are presented. Different with the results shown in Fig.7, in Fig.8, with the increasing of the iteration time, the network overhead decreases before reaching the NE point; when the network reaches NE, the network overhead keeps constant. This result demonstrates that the proposed algorithm is effective on reducing network computation overhead. For different network densities, the more users in the network, the higher computation overhead is. This is because the more users, the more serious inter-cell interference is, which leads to high network computation overhead. However, due to the game between different users, the increasing of the computation overhead with the increasing of the network density become slowly. This means that the algorithm proposed in this paper is effective on improving the performance of MEC system. The Fig.8 also demonstrates that when the computation load increases, the network overhead increases, too. The reason is similar to that shown in Fig.7. The number of iteration times for reaching NE under different network densities is shown in Fig.9. Due to the inter-cell interference, the more users, the more iteration times are needed. The increasing is near a linear, which means that the algorithm can converge in a fast manner.

## VI. CONCLUSION

In this paper, we proposed a game theory based jointing offloading decision and resource allocation algorithm for multi-user MEC. We prove that this game is an exact potential game and the NE of this game exists and is unique. The convergence and the computation complexity of this algorithm are investigated. Moreover: 1) we investigate the PoA of this algorithm and conclude that the interference from the interference users has main effection on the performance of MEC, which is consistent with the conclusion in Corollary 4; 2) we prove that the NE of this game is Pareto efficiency and also the global optimal solution shown in (11). The simulation results shou the effectiveness of this algorithm on improving the performance of MEC.